\documentclass[12pt]{article}

\usepackage{graphicx,amssymb,url,mathrsfs, amsmath}
\usepackage{eucal}
\usepackage{wrapfig}
\usepackage{boxedminipage}
\usepackage{setspace}
\usepackage{subfigure}

\def\a  {\alpha}       \def\b  {\beta}         
       \def\d  {\delta}        \def\D  {\Delta}
     \def\ve {\varepsilon}   \def\k  {\kappa}
\def\l  {\lambda}             
                  
\def\t  {\tau}                 
   \def\w  {\omega}

\newcommand{\cala}{\mbox{${\cal A}$}} \newcommand{\calb}{\mbox{${\cal B}$}}

\newcommand{\calo}{\mbox{${\cal O}$}} 
 
 \newcommand{\calt}{\mbox{${\cal T}$}}

\def\IR{{\hbox{{\rm I}\kern-.2em\hbox{\rm R}}}}
\def\IB{{\hbox{{\rm I}\kern-.2em\hbox{\rm B}}}}
\def\IN{{\hbox{{\rm I}\kern-.2em\hbox{\rm N}}}}
\def\IC{\,\,{\hbox{{\rm I}\kern-.59em\hbox{\bf C}}}}
\def\IZ{{\hbox{{\rm Z}\kern-.4em\hbox{\rm Z}}}}
\def\IP{{\hbox{{\rm I}\kern-.2em\hbox{\rm P}}}}
\def\IH{{\hbox{{\rm I}\kern-.4em\hbox{\rm H}}}}
\def\ID{{\hbox{{\rm I}\kern-.2em\hbox{\rm D}}}}

\def\be{\begin{equation}}
\def\ee{\end{equation}}
\def\ba{\begin{eqnarray}}
\def\ea{\end{eqnarray}}

\def\half{\frac{1}{2}}

\def\fourth{\frac{1}{4}}
\newcommand{\inv}[1]{\frac{1}{#1}}

\def\ra{\rightarrow}  

\newcommand{\dint}[2]{\int_{#1}^{#2}\!\!}

\newcommand{\ud}{\mbox{${\mathrm{d}}$}}
\newcommand{\del}[1]{\partial_{#1}}

\newcommand{\brac}[1]{\langle #1 \rangle}

\renewcommand{\Im}{{ \rm Im}}

\def\setl{\setlength\arraycolsep{2pt}}

\def\nn{\nonumber}
\def\ea{{\it et al}. }

\newcommand{\nt}{{t}}
\newcommand{\nw}{{\w}}
\newcommand{\nnw}{{{ \mathfrak w}}}
\def\tth{\frac{2}{3}}

\begin{document}

\begin{titlepage}


\begin{center}
{\large \bf Diffusion in an Expanding Plasma using AdS/CFT} \\
\vspace{10mm}
  Keun-Young Kim$^a$, Sang-Jin Sin $^{a,b}$  and Ismail Zahed$^a$\\
\vspace{5mm}
$^a$ {\it Department of Physics and Astronomy, SUNY Stony-Brook, NY 11794}\\
$^b$ {\it Department of Physics, Hanyang University, Seoul 133-791, Korea}\\
\vspace{10mm}
 {\tt \today}
\end{center}

\begin{abstract}
We consider the diffusion of a non-relativistic heavy quark
of fixed mass $M$, in a one-dimensionally expanding and strongly
coupled plasma using the AdS/CFT duality. The Green's function
constructed around a static string embedded in a
background with a moving horizon, is identified with
the noise correlation function in a Langevin approach. The
(electric) noise decorrelation is of order $1/T(\tau)$ while the velocity
de-correlation is of order $MD(\tau)/T(\tau)$.
For $MD>1$, the diffusion regime is segregated and the energy loss is Langevin-like.
The time dependent diffusion constant $D(\tau)$ asymptotes its adiabatic limit $2/\pi\sqrt{\lambda} T(\tau)$ when $\tau/\tau_0=(1/3\eta_0\tau_0)^3$ where
$\eta_0$ is the drag coefficient at the initial proper time $\tau_0$.

\end{abstract}

\end{titlepage}

\renewcommand{\thefootnote}{\arabic{footnote}}
\setcounter{footnote}{0}



\section{Introduction}
The quark-gluon plasma (QGP) created in Relativistic Heavy Ion
Collisions at RHIC is believed to be strongly coupled \cite{Shuryak}.
The AdS/CFT correspondence \cite{AdS/CFT} has proven to be a useful
tool for addressing issues of a strongly coupled plasma
albeit in the limit of a large number of colors $N_c$ and strong
gauge coupling $\lambda=g_{YM}^2N_c$. A number of
nonperturbative properties in linear response have been
recently addressed via the gravity dual calculation in AdS$_5$ black
hole geometry~\cite{Policastro}. The transport results bear some relevance
to the QGP plasma at RHIC. Much less is known  about the time-dependent evolution
of the strongly coupled plasma. There have been suggestions that the fireball
in Relativistic Heavy Ion Collision (RHIC) can be explained from a dual gravity
point of view~\cite{SSZ,Nastase}.

In order to model the expanding plasma in a gravity set up, the use of
a moving black hole was suggested in~\cite{SSZ}. In \cite{Janik1, Janik2}
(hereon refer to as JP), it was shown that the moving horizon black hole
geometry can be generated by assuming an asymptotycally expanding perfect
fluid in a boost invariant setting. Using conformal invariance and
energy-momentum conservation together with holographic renormalization
\cite{Haro}, JP constructed the bulk geometry from the perfect fluid
boundary data. This metric was extended to the case of shear viscosity
\cite{Sin,Janik3}, R-charge \cite{Bak} and to a 3 dimensional non-isotropic
setting \cite{SNK}. An exact background with isotropic expansion
was worked out in \cite{Gubser3}.

In this paper we study the diffusion of a non-relativistic heavy
quarks in one dimensionally expanding plasma using the JP metric.
Heavy quark diffusion have been studied in a static black hole
background in various ways
\cite{SZ,Herzog1,Liu,Gubser1,Teaney1,Gubser2,Teaney2}. Here, we
follow the suggestion in \cite{Teaney1} in the static case, by
analyzing the momentum fluctuations of a heavy quark in an
expanding plasma, and use it to estimate the drag force and
diffusion rate. We will use a generalized Langevin equation to
assess the diffusion rate. Unlike \cite{Gubser2} we suggest that
the green function of string fluctuation should be identified
with the correlator of the fluctuation force rather than the
total force in a Langevin approach.

The basic object of this procedure is the gauge-invariant
electric-electric force decorrelation, which we will calculate
using the AdS/CFT duality. The field theory dual (operator) to a
quark displacement($\xi(t)|_{u=0} $) is the (gauge invariant)
colored force acting on a heavy quark
\cite{Teaney1,Gubser2,Teaney2}. According to  the AdS/CFT
correspondence the generating function in field theory should be
related  with the classical action  by
\be \langle \exp({i\int F(t)\xi(t) })\rangle=\exp(iS_{cl}[\xi])\ ,
\ee  whose second derivative gives us the symmetrized Wightman
function
\begin{eqnarray}
G(\nt_1,\nt_2) \equiv \half \brac{  F(\nt_1)F(\nt_2) +
F(\nt_2)F(\nt_1) } \ .
\end{eqnarray}
It is related to the retarded Green's function~\cite{HS}
\begin{eqnarray}
G(\nw) = -\mathrm{coth}\frac{ \nw}{2T_0} \Im G_R(\nw)\ ,
\label{Wightman00}
\end{eqnarray}
in momentum space. We will calculate $G_R(\nw)$ from the
Nambu-Goto action of the fluctuating string in the JP metric
following the way proposed in \cite{Teaney1} (section 2). We note
that this decorrelator applies both to heavy and light
fundamental quarks, therefore the decorrelation time is a measure
of how the gluon rescattering against external fundamental probes
decorrelate.

In section 2 we start with the JP metric in \cite{Janik1} and
derive equations of motion of the transverse string fluctuation
of a heavy quark. We also calculate the retarded Green's
function, which is translated to an electric force-force
correlator (or more precisely decorrelator) in section 3. In
section 4, we compute the momentum fluctuation(broadening) of a
heavy quark. By arguing that a generalized Langevin equation
applies for the case,we deduce the heavy quark diffusion property
in the expanding and cooling plasma. Our conclusions are in
section 5.

\section{String fluctuation in Janik-Peschanski metric}

We consider a heavy particle moving in an expanding plasma. Let
$t_F=1/T$ be the colored electric force decorrelation time (see
below) and let $t_D=MD/T$ be the diffusion time of a particle
with diffusion constant $D$ and mass $M$. For times $t_F<t<t_D$ o
r $MD>1$, the force decorrelation can be treated as
instantaneous. Thus, the velocity decorrelation of a heavy quark
in a strongly coupled plasma can be followed generically by a
Langevin description. For an expanding plasma the description
involves a generalized Langevin of the form

\be
\frac{dp}{dt}+\eta(t)p(t)=F(t),
\ee
with both the drag coefficient $\eta(t)$ and the diffusion constant $D(t)$
time dependent. The idea here is that the force-force decorrelator can be calculated
from the microscopic physics and the drag coefficient $\eta(t)$ is related to the
force-force decorrelator by a non-equilibrium relation \cite{Kubo,Reif}.

For the drag coefficient, we need to calculate the correlators in a frame
which moves with the particle. So it is enough to consider a particle moving with the
expanding plasma. For a one dimensional expansion, such a comoving frame was introduced
by Hwa \cite{Hwa} and  by Bjorken \cite{Bjorken} in terms of  the rapidity $y$ and  the
proper  time $\t$ of the comoving coordinate system. $\t$ and $y$ are related to the
time coordinate $x^0$, and logitudinal coordinate $x^3$, through $ x^0 =
\t\ {\rm cosh}\ y \ $ and $ \ x^3 = \t\ {\rm sinh} \ y \ $.
The Minkowski metric in terms of $\t,y$ can be written
as
\be ds^2=-d\t^2+\t^2dy^2+dx_\perp^2,
\ee
where $x_\perp = \{x^1,x^2\}$ are the transverse coordinates.

The gravity dual of the Hwa-Bjorken system  was worked out in \cite{Janik1,Janik2} and the metric
can be written in the following suggestive form
\begin{eqnarray}
\label{Metric.0} \ud s^2=\frac{R^2}{z^2} \left[- \frac{\left(
1-v^4\right)^2}{(1+v^4)} \ud\tau^2+ \left( {\textstyle
1+v^4}\right) (\tau^2 \ud y^2 +\ud x_\perp^2) + \ud z^2 \right],
\end{eqnarray}
where   $z$ is the fifth coordinate in AdS$_5$ and $v$ is a scaling variable defined by
\begin{eqnarray}\label{v}
v \equiv \frac{z}{(\t/\t_0)^{\frac{1}{3}}}\ \ve^{\fourth}\ , \qquad \ve
\equiv \frac{1}{4}(\pi T_0)^4  \ , 
\end{eqnarray}
Since the horizon is located at $v=1$ or $z\sim \t^{1/3}$,  the black hole horizon may be considered
as moving away from the boundary. We remark that the solution is valid only for
late times (asymptotic solution).

The metric (\ref{Metric.0}) is written in Fefferman-Graham coordinate $z$. It is useful to
express it in a more canonical form, that will prove usefull for a paralell calculation with
the static case. Indeed, by introducing the coordinate change
\begin{eqnarray}\label{u}
u(z,\t) \equiv \frac{2v^2}{1+v^4} \ ,
\end{eqnarray}
\noindent the metric (\ref{Metric.0}) is now
\begin{eqnarray}
\label{Metric.1}
\ud s^2 =&&  \frac{\pi^2 T_0^2 R^2 }{u(\t/\t_0)^{2/3}}\left[ - f(u) \ud\tau^2 +
\tau^2 \ud y^2 +\ud x_\perp^2)\right] +\frac{R^2}{4 f(u) } \frac{\ud u^2}{u^2} \nn \\
&&  +\frac{R^2}{9}\t^{-2} \ud\t^2 -
\frac{R^2}{3}\frac{\t^{-1}}{u\sqrt{f(u)}} \ud \t \ud u \ ,
\end{eqnarray}
with $f=1-u^2$.
We may ingnore the last two terms since the perfect fluid
geometry is valid only in the scaling limit $\t \ra \infty$ and
$v,u \ra \mathrm{constant}$. A further transformation through,
\begin{eqnarray}\label{t.tau}
\nt/t_0 \equiv \frac{3}{2} (\t/\t_0)^{\tth}  \ ,
\end{eqnarray}
yields
\begin{eqnarray}
\label{Metric.2} \ud s^2 = \frac{\pi^2 T_0^2 R^2}{u} \left[- f(u)
\ud \nt^2 + \frac{4}{9} \nt^2 \ud y^2 +
\frac{3}{2}\frac{t_0}{\nt}\ud x_\perp^2)\right] + \frac{R^2}{4
f(u) } \frac{\ud u^2}{u^2}\ .
\end{eqnarray}
This form is  similar to the canonical black hole metic except for the
time dependence of the spacial parts. Also, $t$ and $u$ are not the
same variables as in the static case. They are related to the original
variables, $x^0$ and $z$ through
(\ref{v}), (\ref{u}) and (\ref{t.tau}). In this transformed metric
(\ref{Metric.2}), the black hole horizon is no longer moving away in
the fifth direction but is expanding in the y
direction and contracting in the transverse direction as time goes on.

In the background (\ref{Metric.2}), let us consider the small
string fluctuations in the transverse direction $x_1$,
\begin{eqnarray}
\d X^1 = \xi(\nt, u) \ .
\end{eqnarray}
The relevant Nambu-Goto action is
\begin{eqnarray} \label{String.action}
S = \frac{T_0 \sqrt{\l}}{8 } \dint{0}{\infty} \ud\nt
\dint{-\infty}{\infty} \ud u \left(\frac{3t_0}{2\nt}\right)
\left[
 \frac{(\partial_{\nt} \xi)^2}{u^{\tth}f(u)} - \frac{4f(u)\pi^2 T_0^2}{ u^{\half}}(\partial_u\xi)^2
 \right] \ ,
\end{eqnarray}
where $\sqrt{\l} = \frac{R^2}{\a'}$ after subtracting the
unperturbed string action. Notice that the action
(\ref{String.action}) is the same as the one in the static black hole
metric \cite{Teaney1} except for an overall factor of $\left(\frac{3t_0}{2\nt}\right)$.
The latter stems from the metric component $g_{x^1x^1}$ of (\ref{String.action}), when we
evaluate the induced metric in the Nambu-Goto action.
The equation of motion for $\xi(\nt,u)$ is
\begin{eqnarray}\label{String.eom}
\del{\nt}^2\xi - \inv{\nt}\del{\nt}\xi + 2\pi^4 T_0^4
f(u)(1+3u^2)\del{u}\xi - 4\pi^4 T_0^4 uf(u)^2\del{u}^2\xi= 0 \ .
\end{eqnarray}

To solve (\ref{String.eom}) we define a Fourier-like transform
\begin{eqnarray}\label{Hankel}
\xi(\nt,u) &\equiv& \dint{-\infty}{\infty} \frac{\ud\nw}{2\pi} \
\sqrt{\frac{i \pi \nw}{2}}\ \nt H^{(2)}_1(\nw\nt) \Psi_\nw(u)
\tilde{\xi}_0(\nw) \ ,
\end{eqnarray}
where  $H^{(2)}_1(\nw\nt)$ is a
Hankel function of the second kind, and $\Psi_\nw(u)$ is
normalized such that $\Psi_\nw(0) = 1$.
$t H^{(2)}_1(\nw\nt)$
is chosen to satisfy the time part of the equation of motion
(\ref{String.eom}) with the correct boundary condition (see below).
Notice that we  have extended the region of $t$ from ($0,\infty$) to
($-\infty,\infty$).

To proceed, we assume the following `completeness relation'
\begin{eqnarray}\label{Completeness}
- \frac{1}{4}   \dint{-\infty}{\infty} \ud t \ \nt
H^{(2)}_1(\nw\nt)H^{(2)}_1(-\nw'\nt) \simeq \inv{\nw}\d(\nw -
\nw').
\end{eqnarray}
which will be understood for small $\omega$ ~\footnote{The  usual
completeness of the Hankel transform is in terms of Bessel
functions: $\dint{0}{\infty} \ud t \ \nt
J_\nu(\nw\nt)J_\nu(\nw'\nt) = \inv{\nw}\d(\nw - \nw')$, whose
origin is the asymptotic form of a Bessel function as an
exponential function over $\sqrt{ t}$. There is not true
completeness for the Hankel function $H_{(1,2)}$ due to the
singularity near zero. The use the completeness of the Hankel
function is justified for small $\omega$ or large times since:
1)the dominant integral contribution is coming from the large
time region; 2)the JP background is justified asymptotically. The
use of the Hankel function instead of a Bessel function is needed
to match the incoming bioundary condition below.}. While this
approximation blurs the rigor of the arguments to follow, it
should nevertheless provide us with an insightful understanding
of the time scale  involved in the relaxation of the diffusion
process. These time scales  are paramount to our understanding of
the approach to equilibrium of a strongly coupled quark-gluon
plasma such as the one at RHIC.

After separating the time part, the equation for $\Psi_\nw(u)$
now reads
\begin{eqnarray}\label{Psi.eom}
\del{u}^2\Psi_\nw (u) - \frac{3u^2 +
1}{2uf(u)}\del{u}\Psi_\nw(u) + \frac{\nnw
^2}{4uf(u)^2}\Psi_\nw(u) = 0 \ .
\end{eqnarray}
where $\nnw \equiv \frac{\nw}{\pi T_0}$.  Notice that (\ref{Psi.eom}) is of the same form as the
one in the static black hole metric \cite{Teaney1}. Near the horizon the solution behaves
as
\begin{eqnarray}
\Psi_{\nw} \sim (1-u)^{\pm i \nnw/ 4} \ ,
\end{eqnarray}
and  the $minus$ choice corresponds to the infalling boundary
condition. Inserting (\ref{Hankel}) into the action
(\ref{String.action}) yields the reduced boundary action
\begin{eqnarray}\label{Action:omega1}
&&S_\mathrm{boundary} = \frac{3 \pi^2 \sqrt{\lambda}T_0^3 t_0}{4
} \int \ud\nt \
\frac{f(u)}{\sqrt{u}\nt} \xi(\nt, u) \del{u} \xi(\nt,u)|_{u=0}^{u=1} \nn \\
&&\quad = \int \frac{\ud\nw}{2\pi} \ \tilde{\xi}_0(-\nw) \left[
\left(\frac{3 \pi^2 \sqrt{\lambda}T_0^3 t_0}{4 }\right)
\frac{f(u)}{\sqrt{u}} \Psi_{-\nw}(u) \del{u}
\Psi_\nw(u)\right]_{u=0}^{u=1} \tilde{\xi}_0(\nw) \ ,
\end{eqnarray}
where we used (\ref{Completeness}). Following \cite{Son} we
identify the retarded Green's function, $G_R(\nw)$, as
\begin{eqnarray}\label{Retarded}
G_R(\nw) \equiv \left[-\frac{3 \pi^2 \sqrt{\lambda}T_0^3 t_0
}{2}\right] \Big[\frac{f(u)}{\sqrt{u}} \Psi_{-\nw}(u) \del{u}
\Psi_\nw(u)\Big]_{u=0} \ .
\end{eqnarray}
\section{Electric Force-Force decorrelation}

$G_R(\nw)$ can be calculated analytically only in two limits:
$\nw \ra 0$ and $\nw \ra \infty$.
In the small $\nw $ limit,  we may expand the
solution in terms of $\nnw$ and solve (\ref{Psi.eom}) order by
order with the incoming boundary condition,
\begin{eqnarray}
\Psi_\nw = (1-u)^{-i\nnw/4}\left[1 +
\frac{i\nnw}{2}\left(-\tan^{-1}\sqrt{u}+ \ln(1+\sqrt{u} )\right)
 \right]+\calo(\nnw^2)\ ,
\end{eqnarray}
which gives
\begin{eqnarray}\label{Smallw}
\lim_{\nw \ra 0} \left(\pi \mathrm{coth}\frac{\pi
\nnw}{2}\right) \mathrm{Im} \Big[\frac{f(u)}{\sqrt{u}}
\Psi_{-\nw}(u) \del{u} \Psi_\nw(u)\Big]_{u=0} \ra 1 \ .
\end{eqnarray}
In the large $\nw$ limit,
we can use the WKB approximation, which yields
\begin{eqnarray}
\lim_{\nw \ra 0} \left(\pi \mathrm{coth}\frac{\pi
\nnw}{2}\right) \mathrm{Im} \Big[\frac{f(u)}{\sqrt{u}}
\Psi_{-\nw}(u) \del{u} \Psi_\nw(u)\Big]_{u=0} \ra \frac{ \pi
|\nnw|^3}{2} \ ,
\end{eqnarray}
in agreement with the zero temperature result \cite{Gubser2}.
See Appendix A for more details.

For general $\nw$,  we have to resort to
numerical methods. We should perform the numerical computation and
compare it with the analytic result in the small and large $\nw$ limits
obtained above. The strategy is as follows. \footnote{Similar
calculations have been done in \cite{Teaney3,Kovtun,Gubser2} in other  models.
Our numerical integration is different.}
First we find  two independent  solutions near the horizon ($u \sim 1$)
\begin{eqnarray}\label{Psi.horizon}
\Psi^H_{\nw,in} &\equiv& (1-u)^{-i\nnw/4} \left[1-(1-u)\left(\frac{i\nnw^2}{8i + 4\nnw}\right)  \right]  +\cdots\ ,  \\
\Psi^H_{\nw,out} &\equiv& (\Psi^H_{\nw,i})^*  .\
\end{eqnarray}
Here  $\Psi^H_{\nw,in}$ is the infalling solution and its complex
conjugate is the outgoing solution. Notice that
these solutions  are valid  for all $\nnw$. Near the
boundary ($u\sim 0$), there are two independent solutions
\begin{eqnarray}\label{Psi.boundary0}
\Psi^B_{\nw,0} &\equiv& u^{3/2} - \frac{\nnw^2}{10} u^{5/2} +
\left(\frac{3}{7}+\frac{\nnw^4}{280}\right)u^{7/2} +\cdots\ ,  \\
\label{Psi.boundary1} \Psi^B_{\nw,1} &\equiv& 1 +
\frac{\nnw^2}{2} u - \frac{\nnw^4}{8}u^2 +
\left(\frac{\nnw^2}{9}+\frac{\nnw^6}{144}\right)u^{3} +\cdots \ .
\end{eqnarray}
Notice that $\Psi^B_{\nw,0}$ vanishes, while $\Psi^B_{\nw,1}$ goes to
unity near the boundary and both solutions are real. For the
retarded Green's function, we need the wave function near zero
{\it satisfying infalling boundary condition} at the horizon.
For this, we take the  near-horizon wave-function
(\ref{Psi.horizon}) with the correct boundary condition as the initial data
and numerically integrate it from the horizon  to  the boundary  using (\ref{Psi.eom}).
The solution is expressed as a linear sum of boundary basis $\Psi^B_{\nw,0}$ and $\Psi^B_{\nw,1}$
\begin{eqnarray}\label{numint}
\Psi^H_{\nw,in}(u ) \
\longrightarrow^{\!\!\!\!\!\!\!\!\!\!\!\!\!(\ref{Psi.eom})} \
\cala \Psi^B_{\nw,1}(u ) + \calb
\Psi^B_{\nw,0}(u ) \ .
\end{eqnarray}
where $\cala$ and $\calb$ are complex numbers determined numerically.
Notice that the right hand side goes to $\cala $ at the boundary while we
have to normalize $\Psi$ such that it goes to 1 at $u=0$.  Therefore the
correctly normalized wave function with correct boundary conditions  is
$\Psi_{\nw}=\cala^{-1}\Psi^H_{\nw,in}(u )$:
\begin{eqnarray}\label{Psi.sol}
 \Psi_{\nw}(u)= \Psi^B_{\nw,1}(u ) +  \frac{\calb}{\cala}
\Psi^B_{\nw,0}(u),
\end{eqnarray}
which readily yields
\begin{eqnarray}
\Im \Big[\frac{f(u)}{\sqrt{u}} \Psi_{-\nw}(u) \del{u}
\Psi_\nw (u)\Big]_{u=0} = \frac{3}{2} ~\Im \tilde{\calb}  \ ,
\end{eqnarray}
with  $\tilde{\calb}=\frac{\calb}{\cala}$. Now the Wightman
function  $G(\nw)$ (\ref{Wightman00}) is given by
\begin{eqnarray}\label{Gnnw}
G(\nw) = \left[\frac{3 \pi \sqrt{\lambda}T_0^3\t_0}{2}\right]
\left(\pi \mathrm{coth}\frac{\nw}{2T_0}\right)\left(
\frac{3}{2}~\Im \tilde{\calb}(\nw)\right) \ .
\end{eqnarray}

To complete the numerical calculation, we note that while $\cala$
is easily accessible numerically in (\ref{numint}), $\calb$ is
not. To resolve this problem, we use the following method. First,
by taking the imaginary part of (\ref{Psi.sol}) we get
\be
\Im\tilde{\calb}= \left[ \frac{ {\cala}^{-1} \Psi^H_{\nw,in}(u)}{\Psi^B_{\nw,0}(u)} \right] ,
\ee
and then we evaluate it at any  point, say,  $u=1$.
The only remaining part is the value of $\Psi^B(u)$ at $u=1$, for which we need to
numerically integrate from the boundary to the horizon.
We denote the value determined by this procedure by $\Psi^B_{\nw,0}(u\longrightarrow^
{\!\!\!\!\!\!\!\!\!\!\!\!\!(\ref{Psi.eom})}1)$. Notice $\cala$ is given
before by ${\Psi^H_{\nw,in}(u
\longrightarrow^{\!\!\!\!\!\!\!\!\!\!\!\!\!(\ref{Psi.eom})}
0)}$. Therefore we get the numerical recipe:
\begin{eqnarray}
  \Im\tilde{\calb} = \Im \left[
\frac{\Psi^H_{\nw,in}(u=1-\epsilon)}{\Psi^H_{\nw,in}(u
\longrightarrow^{\!\!\!\!\!\!\!\!\!\!\!\!\!(\ref{Psi.eom})}
0)  \cdot
\Psi^B_{\nw,0}(u\longrightarrow^{\!\!\!\!\!\!\!\!\!\!\!\!\!(\ref{Psi.eom})}1)
 }\right]  \ ,
\end{eqnarray}
 where we take $\epsilon=10^{-6}$.

\begin{figure}[]
  \begin{center}
 \subfigure[] { \includegraphics[width=8cm]{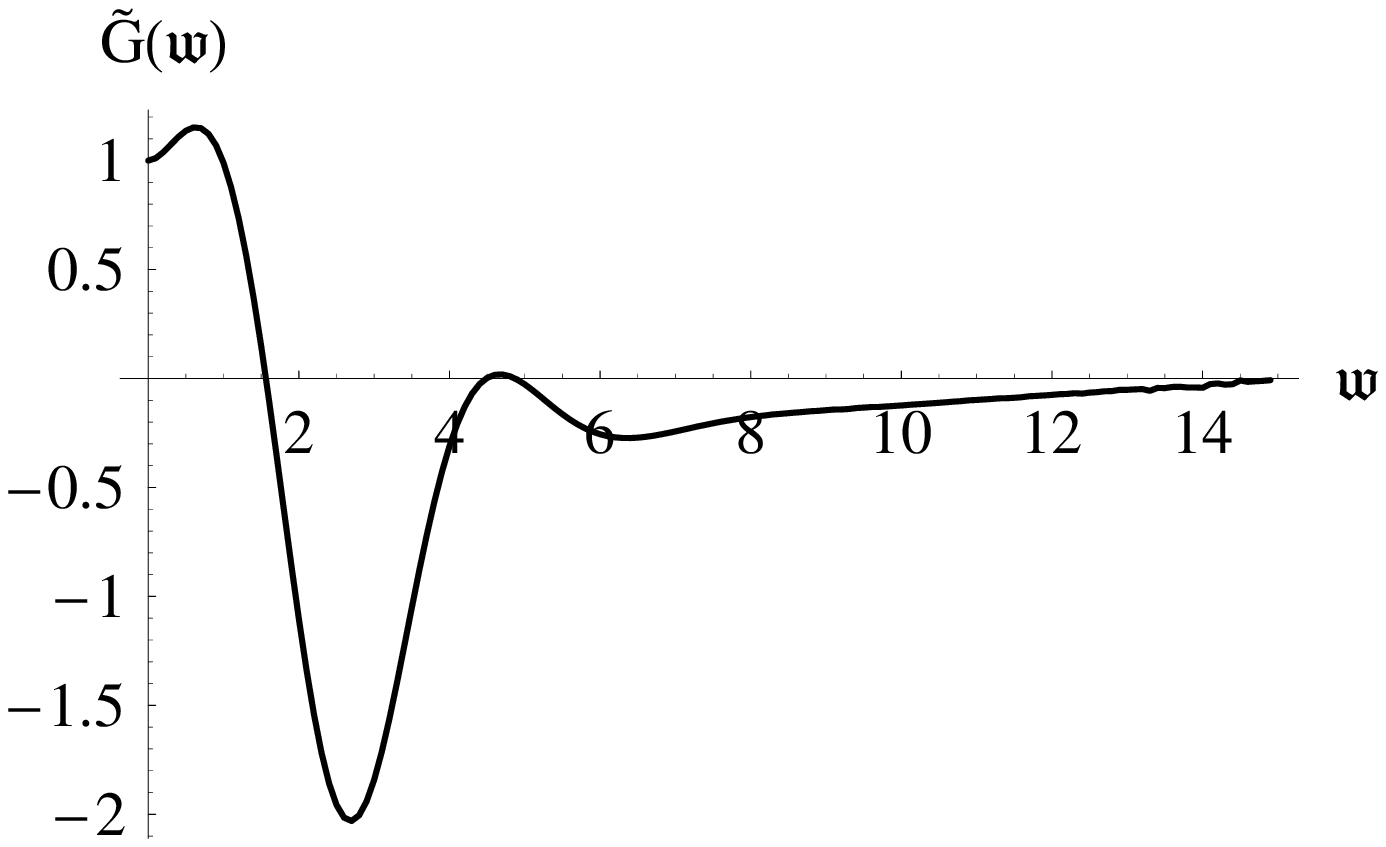}}
 \subfigure[] { \includegraphics[width=8cm]{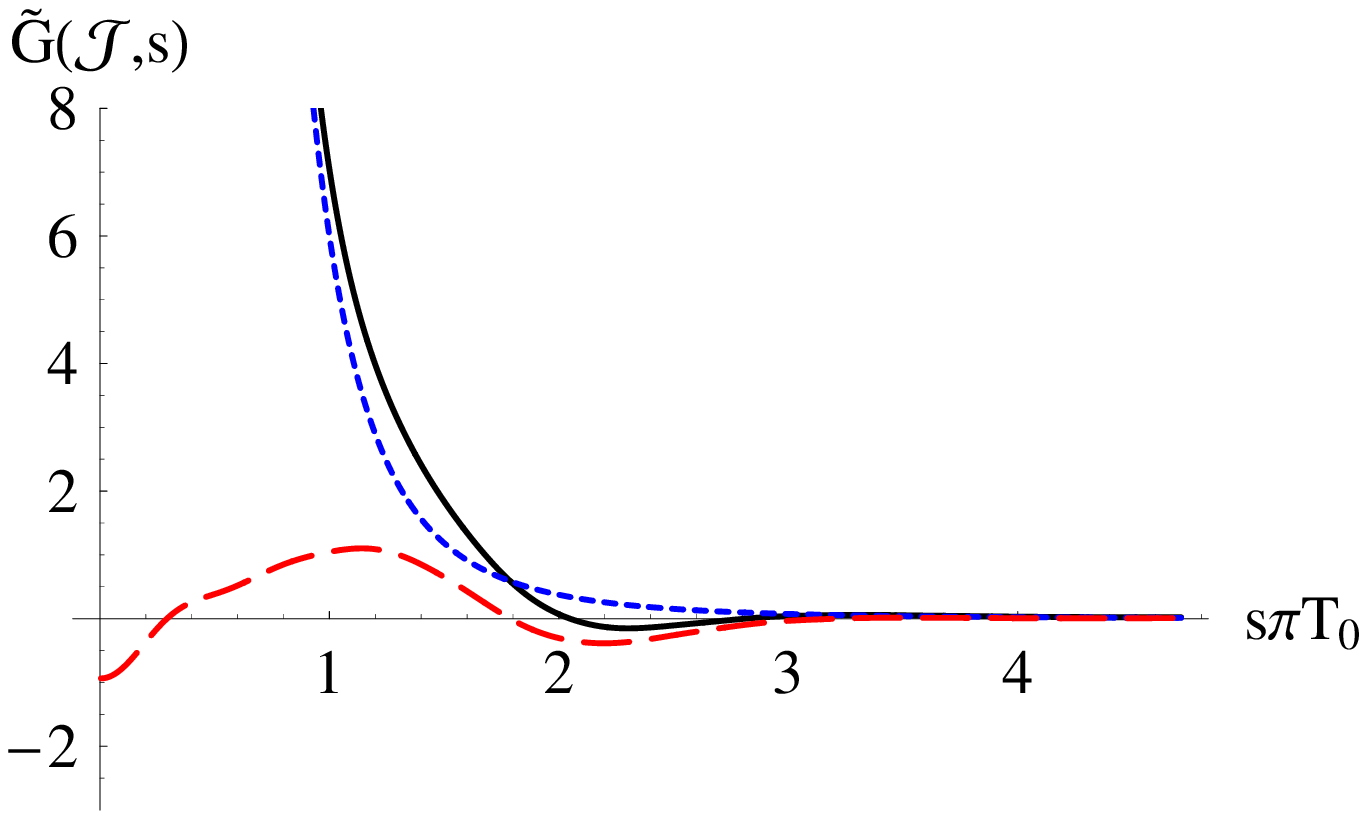}}
  \caption{Force-Force decorrelator: (a) $\displaystyle{\widetilde{G}(\nnw) =
  {G(\nnw)-\frac{\pi}{2}|\nnw^3| \over \left[\frac{3 \pi \sqrt{\lambda}T_0^3\t_0}{2}\right]}}
  $
  (b) $\displaystyle{\widetilde{G}(\calt ,s) =
  {G(\calt,s) \over \left[\frac{3 \pi^2 \sqrt{\lambda}T_0^4\t_0}{2 \calt}\right]}}
  $
  [The long dashed red line: discrete Fourier transform  of (a). The short dashed blue line: the
divergent contribution alone. The solid line: the total result.]}
  \label{ForceCorrelator}
  \end{center}
\end{figure}

The result for (\ref{Gnnw}) is plotted
in~Fig.\ref{ForceCorrelator}a, with the large $\nnw$ asymptotic
$\frac{\pi|\nnw|^3}{2}$ subtracted. There is a good agreement
asymptotically.
The numerical results in time can be obtained by using the
inverse transformation of (\ref{Completeness})~\footnote{
The same comments footnoted after~(\ref{Completeness}) apply.}
\begin{eqnarray}
G(\nt_1,\nt_2)  = -\inv{4} \dint{-\infty}{\infty} d\nw \nw
 H^{(2)}_1(\nw \nt_1 )
H^{(2)}_1(-\nw \nt_2) G(\nw) \ .
\end{eqnarray}
The time-dependence of the problem excludes time-translation
invariance. In terms of the relative and CM coordinates
\begin{eqnarray}
s \equiv t_1 - t_2 \ , \quad \calt \equiv \frac{t_1 + t_2}{2} \ ,
\end{eqnarray}
the force-force  decorrelator is
\begin{eqnarray}
G(\calt, s)&\approx&
 \frac{1 }{\calt} \frac{1}{\sqrt{1-\frac{s^2}{4\calt^2}}} \int \frac{d\nw}{2\pi} \
 e^{-i\nw s } G(\nw) \ ,
\end{eqnarray}
for $t_1,t_2 \gg 1$, which is an ordinary Fourier transform in the
relative time. For $ s \ll \calt$ the asymptote $G(\nw) \sim
|\nnw|^3$, yields
\be \int e^{-i\nw s}|\nw|^3 \sim \inv{s^4} \sim \inv{|t_1-t_2|^4}.
\label{xx} \ee We note that this is a good approximation of the
Hankel transformation (\ref{Hankel}) {\it for large $t_i$}. The
result (\ref{xx}) is expected from the conformal dimension which
is 4 of the force-force decorrelator. Similary for  $ s \sim
\calt$
\begin{eqnarray}
G(\calt, s) \sim \frac{1}{ \calt^4 \sqrt{\calt^2 - s^2/4 }} \ ,
\end{eqnarray}
but it is not reliable due to the nature of our approximation
($t_1,t_2 \gg 1$). In Fig.\ref{ForceCorrelator}a, we show the
regular part of $G(\nw)$. In Fig.\ref{ForceCorrelator}b, we show
its discrete Fourier transform in dashed red versus $s\pi T_0$.
The short dashed blue line is the divergent contribution alone
which is dominant at small relative times. The solid line is the
total result. The decorrelation time follows readily from the long
dashed red curve as
\begin{eqnarray}
t_F\sim \frac 2{\pi T_0}\ .
\end{eqnarray}
This result is important as it indicates that from dual AdS/CFT
{\cal all} electric-electric forces applied to either {\it heavy}
or {\it light} fundamental probes decorrelate on a {\it short}
time scale of the order of $2/\pi T_0$ in the static but strongly
coupled QGP. We also note that the decorrelation curve in long
dashed-red is stronger than exponential. This time compares
favorably with the time read from the lowest quasi-normal mode
$\nnw_1^{qn}$ associated to string fluctuations (see Appendix B)
\be \nnw_1^{qn}\approx 2.69 - 2.29i.\ee
This yields a decorrelation time of order $0.44/T_0$ which is
comparable to our $0.64/T_0$ numerical estimate~\footnote{The
same value was reported in \cite{Gubser2} for the fluctuation of
the trailing string.}

Using the relation (\ref{t.tau}),
\begin{eqnarray}
t_F=\d t = (\t_0/\t)^{1/3}\d\t \ ,
\end{eqnarray}
this 'static' time translates to a 'moving' time
\begin{eqnarray}\label{Time}
\d\t \sim \frac{(\t/\t_0)^{1/3}}{T_0} \equiv  \inv{T(\t)},
\end{eqnarray}
which is the natural time dependent temperature, $T(\t)$, in agreement
with Bjorken hydrodynamics~\cite{Bjorken,Janik1}.

\section{Momentum fluctuation and Diffusion of a heavy quark}
The short force-force decorrelation time assessed above justifies
the arguments presented in section 2. Again in the time window
$t_F<\Delta t<t_D$, the fluctuations in the momentum of a massive
quark with $MD>1$ read
\begin{eqnarray}
\brac{\D p(\nt)^2} &\equiv& \brac{(p(\nt + \D\nt)- p(\nt))^2} \ .
\end{eqnarray}
For a quark at rest this accounts for its momentum broadening
by thermal quicks. Using (\ref{Smallw}), then
{\setl
\begin{eqnarray}
\brac{\D p(\nt)^2}  &=& \dint{\nt}{\nt + \D \nt} \ud \nt_1
\dint{\nt}{\nt+\D \nt}\ud \nt_2 \brac{F(\nt_1)F(\nt_2)} \approx
 \dint{\nt}{\nt+\D\nt} \ud \calt \dint{-\infty}{\infty}\ud s G(\calt,s)  \nn \\
&=&  \ln \left( 1 + \frac{\D\nt}{t} \right) \left[\frac{3
\pi\sqrt{\lambda}T_0^3 t_0}{2}\right] \lim_{\nnw \ra 0} \left(\pi
\mathrm{coth}\frac{\pi \nnw}{2}\right) \mathrm{Im}
\Big[\frac{f(u)}{\sqrt{u}} \Psi_{-\nw}(u) \del{u}
\Psi_\nw,(u)\Big]_{u=0} \nn \\
&\approx& \frac{3}{2}\pi \sqrt{\lambda}T_0^3 t_0 \frac{\D
\nt}{\nt} \ ,
\end{eqnarray}
}where we have used the fact that $G(\calt,s)$ is well localized.
From (\ref{t.tau}), the time dependent momentum transfer is
\begin{eqnarray}
\brac{ \D p( \t )^2} = \pi \sqrt{\lambda} T_0^3 t_0 \frac{\D
\t}{\t} :=\k(\t){\D \t}\ .
\end{eqnarray}
For Brownian diffusion which is the case here given the short decorrelation
time of the electric force, this amounts to
\begin{eqnarray}\label{kappa}
\k(\t)  =  \frac{ \pi
\sqrt{\lambda}T_0^3}{\t / \t_0} = \pi \sqrt{\l}T^3(\t) \ .
\end{eqnarray}
with $T(\tau)$ defined in (\ref{Time}). $\kappa (\tau)$ is the time-dependent
momentum diffusion constant.

Now consider the diffusion of a nonrelativistic heavy quark in
the medium of which temperature is cooling down adiabatically.
From above it follows that the Langevin equation captures the
essentials of the equilibration in the diffusion regime. Since
the medium is expanding, the appropriate description is given by
\begin{eqnarray}\label{Langevin}
\frac{\ud p(\t)}{\ud \t} = - \eta_D(\t) p(\t) +  {F(\t)} \ ,
\quad \brac{F(\t)} = 0 \ .
\end{eqnarray}
where $\eta(\t)$ is a time-dependent drag coefficient which is
related to $F(\t)$ by
\begin{eqnarray} \label{FD}
\eta(\t) = \frac{1}{2M   T(\t)} \frac{\ud}{\ud \t}\dint{0}{\t}\ud
\t_1 \dint{0}{\t}\ud \t_2 \brac{F(\t_1)F(\t_2)} = \frac{\k(\t)}{2M
  T(\t)} \ .
\end{eqnarray}
One way to derive this is to follow~\cite{Reif}, where all
arguments are done for small time steps $\D\nt$ whereby the
expanding medium is frozen.

Multiplying both sides of (\ref{Langevin}) by $x(\t)$ and taking
the ensemble average of the product one gets an ordinary
differential equation,
\begin{eqnarray}\label{Langevin.1}
 \frac{\ud^2}{\ud \t^2} \brac{{x}^2} +
\eta(\t) \frac{\ud}{\ud \t}\brac{{x}^2} -2  \brac{v(\t)^2} = 0 \ .
\end{eqnarray}
where the only property of $F(\t)$ we need is $\brac{x(\t)F(\t)} =
\brac{x(\t)}\brac{F(\t)}= 0 $, which is a basic assumption of
the Langevin equation. The time-dependent diffusion rate $D(\t)$
reads
\begin{eqnarray}
D(\t) \equiv \half\frac{\ud}{\ud \t}\brac{{x}^2} \ .
\end{eqnarray}
so that (\ref{Langevin.1}) is
\begin{eqnarray}\label{Diffusion}
\dot{D}(\t) + \eta(\t) D(\t) - \brac{v(\t)^2} = 0 \ .
\end{eqnarray}
To solve (\ref{Diffusion}) we need two inputs:
$\eta(\t)$ and $\brac{v(\t)^2}$. $\eta(\t)$ is given by
(\ref{kappa}) and (\ref{FD}),
\begin{eqnarray}\label{Eta}
\eta(\t) = \frac{\pi \sqrt{\l}T(\t)^2}{2 M } \ ,
\end{eqnarray}
which is similar to the static case~\cite{Herzog1,Gubser1,Teaney1}
except for the time dependent temperature. This confirms the adiabatic
nature of the expansion in the case of short force-force decorrelations.
Using the adiabatic form of the equipartition theorem, we have
\begin{eqnarray}\label{V2}
<v(\t)^2> = \frac{T(\t)}{M } \ .
\end{eqnarray}

Using (\ref{Eta}) and (\ref{V2}), we now have
\begin{eqnarray}\label{Diffusion.1}
  \dot{D}(\t) + a\ \t^{-2/3} D(\t) - b\ \t^{-1/3} = 0 ,
\end{eqnarray}
with $a = \eta_0\t_0^{2/3}$ and  $ b ={T_0\t_0^{1/3}}/{M} $. The solution is
\setl{
\begin{eqnarray}
D(\t) &=& \frac{b}{a}\t^{1/3} + D(0) e^{-3a\t^{1/3}}
\ .\nn
\end{eqnarray}
}This result is important as it shows how the diffusion rate for
a quark changes in an expanding medium. At short times it is
$D(0)$ while at large times it asymptotes
\be D(\tau)= \frac{2}{\pi\sqrt{\l}T(\t)}. \label{yy} \ee which is
the result in~\cite{Teaney1} with an adiabatically changing
temperature. In a way, this justifies a posteriori the use of the
`completeness relation' (\ref{Completeness}) for large times
$\calt $. The cross over between short and long times is
exponential and of order $3a\tau^{1/3}=1$ in time. so (\ref{yy})
is reached for $\tau/\tau_0=(1/3\eta_0\tau_0)^3$. At RHIC
$\tau_0\approx 1$ fm so that $\tau/\tau_0\approx 1/\eta_0^3$ in
the Bjorken phase. Although our arguments rely on large times
within the diffusion window (see above) the cross over regime can
still be approached albeit from above.

\section{Conclusion}

We have analyzed the diffusion of a heavy quark in an expanding and
strongly coupled QGP using the AdS/CFT construction. Our arguments
provide some insights to a truly non-equilibrium phenomenon at strong
coupling. Our analysis was restricted to an asymptotically Bjorken
expanding fluid which is dual to the JP metric.

In the comoving frame the time-dependent diffusion problem is mapped
onto a time-independent one, whereby the diffusion is captured in
a retarded Green's function with proper boundary conditions in the
gravity dual space. The Green's function reflects on the electric
nature of the noise in the rest frame of a massive quark. It is
important to note that the noise (force-force) decorrelation
is short with $t_F\sim 2/\pi T_0$ for both ligh and heavy fundamental
quarks. A strongly coupled QGP randomize the electric correlations
very efficiently.

If we denote by $T_D=MD/T$ the diffusion time, for times $t_F<t<t_D$
a diffusion regime for massive quarks open up whereby a generic
and memoriless Langevin description holds. This desciption requires
the knowledge of only two underlying moments of the phase space
distribution: the average shift captured by the drag and the
diffusion constant. Our construction allows the generalization of
these concepts to a time-dependent Langevin description pertinent
for an expanding fluid. We have found that asymptotic diffusion
sets in on a time scale $\tau\sim 1/\eta_0^3$. This estimate
is reached from the generic character of the Langevin description.

Finally and while we have used the construct in~\cite{Policastro}
for the retarded Green's function, it will be of some interest  to check
explicitly this using the arguments presented in~\cite{Herzog2}.

\section{Acknowledgments}

The work of KYK and IZ was supported in part by US-DOE grants
DE-FG02-88ER40388 and DE-FG03-97ER4014. The work of SJS was
 supported by KOSEF Grant  R01-2007-000-10214-0 and
  also by the CQUEST  with grant number R11-2005-021.
\appendix

\section{WKB  for large $\nnw$}

In this appendix we solve (\ref{Psi.eom}) for large $\nnw$.
To use the WKB approximation in~\cite{Son,Teaney3}, we need to
transform (\ref{Psi.eom}) to a Schr\"{o}dinger type equation.
By a change of variable
\begin{eqnarray}\label{phi}
\phi(u) \equiv \frac{\sqrt{1-u^2}}{u^{1/4}} \Psi_\nw(u)\ ,
\end{eqnarray}
(\ref{Psi.eom}) transforms to
\begin{eqnarray}\label{Schrodinger}
\frac{\ud^2 \phi}{\ud u^2} - V \phi = 0 \ , \quad V \equiv -
\frac{-5+18u^2 + 3u^4 + 4u\nnw^2}{16 u^2 (1-u^2)^2 } \ ,
\end{eqnarray}
where $V$ plays the role of a potential. $V$ will be approximated as
follows,
\begin{eqnarray}
u \approx 1 : \qquad &-& \frac{4 + \nnw^2}{16(1-u)^2} \ , \nn \\
\mathrm{WKB} : \qquad &-& \frac{\nnw^2}{4 u (1-u^2)^2} \ , \nn  \\
u \approx 0  : \qquad & &  \frac{5}{16 u^2} - \frac{\nnw^2}{4 u} \ ,
\end{eqnarray}
and the corresponding solutions are
\begin{eqnarray}
u \approx 1 : \qquad &&  (1-u)^{\half - \frac{i\nnw}{4}}\ , \quad (1-u)^{\half +
\frac{i\nnw}{4}} \ , \nn \\
\mathrm{WKB}: \qquad && \sqrt{\frac{2\sqrt{u}(1-u^2)}{\nnw}}\ e^{
i\half (\nnw
\tan^{-1}(\sqrt{u}) + \nnw \tanh^{-1}(\sqrt{u}) )} \ , \nn \\
&&   \sqrt{\frac{2\sqrt{u}(1-u^2)}{\nnw}}\ e^{- i\half (\nnw
\tan^{-1}(\sqrt{u}) + \nnw \tanh^{-1}(\sqrt{u}) )} \ , \nn \\
u \approx 0 :  \qquad &&
\frac{1}{u^{1/4}}e^{i\sqrt{u}\nnw}(1-i\sqrt{u}\nnw)\ , \quad
\frac{1}{u^{1/4}\nnw^3}e^{-i\sqrt{u}\nnw}(1+i\sqrt{u}\nnw) \ .
\end{eqnarray}
Since the solutions are valid only in the limited region we need
to tie them to fulfill the expected boundary conditions.
There are two boundary conditions: One is the incoming boundary
condition near the horizon and the other is the
normalization ($\Psi_\nw, (0) = 1$). The incoming solution near
the horizon is $\Psi_\nw, \sim  (1-u)^{- \frac{i\nnw}{4}}$ and it
corresponds to $ \phi = (1-u)^{\half - \frac{i\nnw}{4}}$. It can
be shown that all the first parts of the solutions are
connected to each other so that the physical solution near $u = 0$ is
\begin{eqnarray}
\Psi_\nw, = e^{i\sqrt{u}\nnw}(1-i\sqrt{u}\nnw) \ ,
\end{eqnarray}
which gives us
\begin{eqnarray}
\lim_{\nnw \ra 0} \mathrm{Im} \Big[\frac{f(u)}{\sqrt{u}}
\Psi_{-\nnw}(u) \del{u} \Psi_\nw,(u)\Big]_{u=0} \ra \frac{
\pi\nnw^3}{2} \ .
\end{eqnarray}
\section{Quasi-normal modes}
In this section we compute the lowest quasi-normal mode following
the method presented in~\cite{Horowitz}. By a change of variable
\begin{eqnarray}
y \equiv 1-u \ ,
\end{eqnarray}
Eq.(\ref{Psi.eom}) can be reduced to the Heun equation:
\begin{eqnarray} \label{Heun2}
\partial_y^2 \psi(y) + \frac{3(1-y)^2+1}{2 y (1-y)(2-y)}\partial_y
\psi(y)+ \frac{\nnw^2}{4 y^2 (1-y) (2-y)^2}\psi(y) = 0 \ .
\end{eqnarray}
where y=0(at horizen) is a regular singular point with
characteristic exponent $\{-i\frac{\nnw}{4},i\frac{\nnw}{4} \}$.
The quasinormal mode is the solution of (\ref{Heun2}) obeying the
incoming boundary condition at the horizen $y=0$ and the
vanishing Dirichlet boundary condition at $y=1$. So we choose
the exponent $-i\frac{\nnw}{4}$ at $y=0$. To match the boundary
condition at $y=1$, it is convenient to transform (\ref{Heun2})
once more by,
\begin{eqnarray}
\psi(y) \equiv y^{-i\frac{\nnw}{4}} (y-2)^{-\frac{\nnw}{4}}
\chi(y) \ .
\end{eqnarray}
Then (\ref{Heun2}) is reduced to the standard form of the Heun
equation.
\begin{eqnarray}
\partial_y^2 \chi(y) +
\left(\frac{\gamma}{z}+\frac{\delta}{z-1}+\frac{\epsilon}{z-2}\right)\partial_y
\chi(y) +\frac{\a\b z - q}{z(z-1)(z-2)}\chi(y) =0 \ ,
\end{eqnarray}
where
\begin{eqnarray}
\a \b := \frac{i \nnw^2}{8} - \frac{\nnw}{8}(1+i) \ , \quad
\gamma := 1-i \frac{\nnw}{2} \ , \\ \nn \delta = -\frac{1}{2} \ ,
\quad \epsilon =1- \frac{\nnw}{2} \ , \quad q := -\frac{
\nnw^2}{8}(2-i) - \frac{\nnw}{4} \ .
\end{eqnarray}
At $y=0$, the local series solution corresponding to the zero
characteristic exponent and normalized to 1 is given by
\begin{eqnarray}\label{series1}
\chi_0(y) = \sum_{n=0}^{\infty} a_n(\nnw) y^n \ ,
\end{eqnarray}
where
\begin{eqnarray}
&& a_0 = 1 \ , \quad a_1 = \frac{q}{2\gamma} \ , \\ \nn && a_{n+2}
+ A_n(\nnw)a_{n+1} + B_n(\nnw) a_n = 0 \quad (n\geq 2) \ , \\
\nn && A_n(\nnw) := - \frac{(n+1)(2\delta + \epsilon +
3(n+\gamma)+ q}{2(n+2)(n+1+\gamma)} \ , \\ \nn && B_n(\nnw) :=
\frac{n^2+n(\gamma + \delta + \epsilon -1) + \a \b
}{2(n+2)(n+1+\gamma)} \ .
\end{eqnarray}
At y=1, we get the boundary value using the series (\ref{series1})
\begin{eqnarray}\label{series2}
\chi_0(1) = \sum_{n=0}^{\infty} a_n(\nnw) \ ,
\end{eqnarray}
To find the quasinormal modes, we need to find the
zeroes of (\ref{series2}) in the complex $\nnw$ plane. This is
done by truncating the series after a large number of terms
\begin{eqnarray}
|\chi_0(1)^N|^2 := |\sum_{n=0}^{N} a_n(\nnw)|^2 = 0 \ .
\end{eqnarray}
We search for the minimum of $|\chi_0(1)^N|^2$, and check that
the minimum value is zero.

\end{document}